\magnification=\magstep1

\vsize=9.2truein
\def\medskip{\vskip 12truept plus 0truept minus 0truept}
\parskip=0truept plus 0truept minus 0truept
\parindent 2em

\hsize=6.5truein
\hoffset=0.00truein
\baselineskip=16truept plus 0 truept minus 0 truept
\overfullrule=0pt

\newcount\notecnt
\notecnt=0
\def\adnote{\advance\notecnt by1}
\def\fnote{\adnote \footnote{$^{[\number\notecnt]}$}}

\pageno=-1

\centerline{\bf Collisional Evolution in the Vulcanoid Region:}
\centerline{\bf Implications for Present-Day Population Constraints}

\bigskip

\smallskip

\centerline{S.~Alan Stern and Daniel D.~Durda}
\centerline{Department of Space Studies}
\centerline{Southwest Research Institute}
\centerline{Boulder, Colorado 80302}

\centerline { }
\centerline { }
\centerline { }
\centerline { }
\centerline { }
\centerline { }
\centerline { }

\centerline { }
\centerline { }
\centerline { }
\centerline { }
\centerline { }

\centerline { }
\centerline { }
\centerline { }
\centerline { }
\centerline { }

\centerline { }
\centerline { }
\centerline { }
\centerline { }

\centerline { }  
\centerline { }  
\centerline { }

\bigskip

\noindent 15 Pages

\noindent 06 Figures

\noindent 01 Table 

\smallskip
\noindent Submitted to: {\it Icarus,} May 1999 

\smallskip
\noindent Revised: October 1999

\vfill
\eject

\noindent {\bf Vulcanoid Collisional Evolution}

\medskip

S.~Alan Stern

Space Studies Department

Southwest Research Institute

1050 Walnut Street, Suite 426 

Boulder, CO 80302 

\medskip
[303] 546-9670 (voice) 

[303] 546-9687 (fax)

\medskip
astern@swri.edu

\vfill
\eject

\pageno=1

\centerline {\bf ABSTRACT}

\medskip \noindent We explore the effects of collisional evolution on putative Vulcanoid
ensembles in the region between 0.06 and 0.21 AU from the Sun, in order to constrain the
probable population density and population structure of this region today.  Dynamical
studies have shown that the Vulcanoid Zone (VZ) could be populated. However, we find
that the frequency and energetics of collisional evolution this close to the Sun,
coupled with the efficient radiation transport of small debris out of this region,
together conspire to create an active and highly intensive collisional environment which
depletes any very significant population of rocky bodies placed in it, unless the bodies
exhibit orbits that are circular to $\sim$10$^{-3}$ or less, or highly lossy mechanical
properties that correspond to a fraction of impact energy significantly less than 10\%
being imparted to ejecta. The most favorable locale for residual bodies to survive in
this region is in highly circular orbits near the outer edge of the dynamically stable
Vulcanoid Zone (i.e., near 0.2 AU), where collisional evolution and radiation transport
of small bodies and debris proceed most slowly.  If the mean random orbital eccentricity
in this region exceeds $\sim$10$^{-3}$, then our work suggests it is unlikely that more
than a few hundred objects with radii larger than 1 km will be found in the entire VZ;
assuming the largest objects have a radius of 30 km, then the total mass of bodies in
the VZ down to 0.1 km radii is likely to be no more than $\sim$10$^{-6}$M$_{\oplus}$,
$<$10$^{-3}$ the mass of the asteroid belt. A 0.01 AU wide ring near the outer stability
boundary of the VZ at 0.2 AU would likely not contain over a few tens of objects with
radii larger than 1 km. Despite the dynamical stability of large objects in this region
(Evans \& Tabachnik 1999), it is plausible that the entire region is virtually empty of
km-scale and larger objects.

\vfill
\eject

\centerline {\bf 1. INTRODUCTION}

\medskip \noindent Over the past 20 years, our understanding of the solar system has grown
dramatically, as evidenced by the detection of a series of heretofore wholly or largely
undetected populations of small bodies.  These include the Kuiper Belt region beyond Neptune, the
population of Centaurs orbiting in the giant planet region, and the near Earth asteroid (NEA)
zone (for additional background on each, see articles in the review volume edited by Rettig \&
Hahn [1996]).  These new populations are revealing valuable insights into both the architecture
of our solar system (and by extension, others), and the nature and origin of small bodies, and
with regard to impact hazards on Earth.

\medskip \noindent Among the few stable dynamical niches which remain largely unexplored today is
the region interior to Mercury's orbit, where a population of small, asteroid-like bodies called
the Vulcanoids has long been hypothesized to reside (e.g., Perrine 1902; see Campins et al.~1996
for a recent review).  This putative reservoir is of interest because it would plausibly contain
a sample of condensed material from the early inner solar system, and because it would bear
relevance to our understanding and the interpretation of Mercury's cratering record, and thus
Mercury's surface chronology.  Owing to the intense thermal conditions and comparatively high
collision velocities characteristic of this region, the Vulcanoid population might also be
expected to contain unique chemical (e.g., ultra-refractory) signatures not seen in more
heliocentrically distant, small body reservoirs.

\medskip \noindent The ``Vulcanoid Zone'' (VZ) extends inward from a stability limit near 0.21
AU, set by orbital eccentricity excitations due to Mercury and the other planets (Leake et
al.~1987, Evans \& Tabachnik 1999; S.~Brooks, priv.~comm 1999).  The VZ is likely to be
effectively bounded on the inside by the combination of thermal conditions and dynamical
transport effects (i.e., Poynting-Robertson (PR) drag and the Yarkovsky effect; e.g., Leake et
al.~1987; Campins et al.~1996).  Even pure Fe bodies with radius r$<$50 km would evaporate under
solar insolation in 4.5 Gyr at or inside 0.06 AU, and pure Fe bodies with radius r$<$1 km would
evaporate under solar insolation in 4.5 Gyr at or inside 0.07 AU (Lebofsky 1975; Campins et
al.~1996).  PR drag extends this limit outward somewhat because it can move a $\rho$=4 gm
cm$^{-3}$, 1 km radius object from 0.08 to 0.07 AU in 4 Gyr, where it would then be evaporated;
the Yarkovsky effect may dominate over PR drag, thereby removing some 1 km-scale primordial
objects from the zone from even greater distances.  Based on these results, we adopt for what
follows an effective inner boundary of the VZ at 0.06 AU, but point out that if a population of
primordial objects were to exist inside 0.1 AU, a steep heliocentric depletion would be expected
to manifest itself inside $\approx$0.08 AU.

\medskip \noindent Unfortunately, despite the fact that the Vulcanoid region is a plausible
dynamical reservoir for small bodies, any Vulcanoid population will be particularly hard to
detect.  This is because the small bodies believed to be there are close to the Sun (in angular
terms),\fnote {From 1 AU, the VZ inner and outer limits correspond to maximum solar elongation
angles of only 4 deg to 12 deg.}  and comparatively faint (i.e., 9$<$V$<$13) compared to the sky
at twilight.  The angular proximity of the VZ to the Sun itself constrains groundbased
visible-wavelength searches to brief windows of difficult, twilight geometry, or alternatively,
to total solar eclipses.  As a result, few VZ searches have been carried out, and those that have
(e.g., Campbell \& Trumpler 1923; Courten 1976; see Campins et al.~1996) exhibited comparatively
shallow limiting magnitudes.  Still, owing to the strongly increased flux of the Sun on the
Vulcanoid region, even these early studies were sensitive to objects with radii down to $\sim$50
km. No objects were discovered.

\medskip \noindent Visible-wavelength searches to date have covered most of the VZ inside 0.25 AU,
but only reached V$\approx$8.5.  As shown in Figure 1, this corresponds to comparatively large
objects with radii of 30 to 50 km at 0.20 AU; objects smaller than this would have escaped
detection, even if present in great numbers.  The most constraining search published to date
worked in daylight conditions to detect the thermal-IR signature of Vulcanoids in the L (3.5
$\mu$m) band (Leake et al.~1987).  This effort reached a magnitude limit of L=5, corresponding to
objects with radii near 3.5 km at 0.21 AU, but covered only 5.8 deg$^2$ of sky, which is $<$5\%
of the available search area.  Owing to the small area of this search, it is not possible to rule
out populations containing a few objects with radii exceeding 25 km, and a some dozens of
objects larger than 5 km in radius; the population of still smaller objects remains almost wholly
unconstrained.\fnote {Searches for IR emission from dust close to the Sun which might result from
recent collisions among Vulcanoids (e.g., Hodapp et al.~1992, MacQueen et al.~1995) have also
yielded negative results.  A study of the zodiacal light using the photometer aboard the Helios
spacecraft (Leinert et al.~1981) never penetrated the region inside 16 deg from the Sun where the
Vulcanoids are expected to reside.}  Together these various observational results imply that the
present-day VZ certainly cannot contain a large population of objects with radii in excess of 3.5
km, and likely contains zero (or perhaps only a handfull) of objects with radii of $>$25 km.

\medskip \noindent This brief summary recapitulates much of what is known about the Vulcanoid
Zone, and demonstrates that an ensemble of small bodies in the size range 1 km to a fews 10s of
km in radius could exist and remain undetected there.  In this report we examine the effects of
collisional evolution on a suite of hypothetical Vulcanoid populations, with the specific
objective of further constraining the extent of any small-body population interior to Mercury's
orbit.

\centerline { }
\centerline { }
\centerline {\bf 2. COLLISION ENERGETICS IN THE VULCANOID ZONE}

\medskip \noindent The consequences of collisions in the Vulcanoid Zone depend on whether the
collisional environment promotes net erosion or net accretion.  As one intuitively expects,
collisions in the VZ are highly energetic, owing to the high Keplerian orbital velocities close
to the Sun.  To illustrate this point, consider an orbiting swarm with mean random inclination
$i$ and eccentricity $e$ in approximate statistical equilibrium, i.e., $\langle i \rangle$=${1
\over 2}$$\langle e \rangle$, the mean encounter speed at infinity,\fnote {i.e., Neglecting
mutual gravitational acceleration due to the binding energy of the impactor and target.}  as a
function of heliocentric semi-major axis $a$, is of scale:

$$
V_{enc} = 180 \langle e \rangle \sqrt{{0.1~{\rm AU}} \over {a}}~~{\rm km~s^{-1}}.  \eqno (1)
$$

\medskip \noindent Even higher encounter speeds would be achieved if $\langle i \rangle$$>$${1
\over 2}$$\langle e \rangle$, as is the case in the asteroid belt.  Still, even for the case in
Eqn.~(1) and $\langle e \rangle$=0.01, $V_{enc}$ is $\sim$2 km s$^{-1}$, over an order of
magnitude higher than the 0.1 km s$^{-1}$ escape speed from a 50 km radius body with density
equal to the Earth's iron core.  As such, even for lossy collisions into mechanically strong
objects in the VZ, one expects collisions to be highly erosive.

\medskip \noindent This result can be further quantified by adopting an analytical formalism
which derives a critical collision velocity, or equivalently, a critical orbital eccentricity
$e^*$, above which impacts eject more mass from the object than the mass of the impactor, and
below which the target body gains mass and thereby grows (Stern 1995).  This critical
eccentricity is a function of several target parameters, including strength, size, and mass.

\medskip \noindent The results of a set of $e^*$ calculations for the Vulcanoid Zone are shown in
Figure 2.  This figure shows that, assuming f$_{KE}$=0.08, objects across the VZ with radii of a
few km or less will suffer erosion by impacts even if the orbital eccentricities of the colliding
objects are as low as 5$\times$10$^{-5}$ to 10$^{-4}$, depending on their mechanical properties.
For both weak and strong mechanical properties, $e^*$ is $\sim$2 times lower at the 0.06 AU VZ
inner-boundary than at the 0.21 AU VZ outer boundary.  Larger objects are more resistive to
erosion owing to their gravitational binding energy, which acts to return low-velocity ejecta.
Still, however, their $e^*$ boundary also occurs at comparatively low orbital eccentricities,
owing primarily to the high Kepler velocities, and therefore the high specific impact energies
inherent in the VZ.  Even the largest objects still marginally permitted by searches, i.e., those
with radii near 50 km, will suffer erosion if their orbital eccentricities are as low as
10$^{-3}$ to 5$\times$10$^{-3}$, depending on their mechanical properties.

\medskip \noindent How do these critical eccentricity results compare to expected eccentricities
in the region?  One worthwhile comparison is obtained by noting that a population of 10 km radius
bodies would mutually excite orbital eccentricities to levels of 10$^{-4}$ in the VZ if there is
no substantial population of still smaller bodies causing dynamical drag.  Mean random
eccentricities of the 10$^{-4}$ level could also have been excited by a former population of
objects in Mercury's feeding zone with masses of 10\% to 20\% of Mercury.  Still larger
eccentricities could have been excited either by large interlopers in the region, or by sweeping
secular resonances, including those generated by solar spin-down (Ward et al.~1976).  Given these
considerations, and the characteristic eccentricity levels seen in the terrestrial planet zone,
we consider it plausible that the mean random eccentricity of VZ orbits could be as low as a
10$^{-4}$ or as high as several times 10$^{-1}$.

\medskip \noindent Given the low critical eccentricity required for erosion in the VZ, and
the plethora of dynamical processes which could have excited the region to random orbital
eccentricities above $e^*$, we find it highly unlikely that the $e^*$ boundary has not
been exceeded; therefore, present-day collisions in the VZ are likely to be erosional.

\medskip \noindent This is not a surprising conclusion.  However, it does imply the interesting
corollary that the conditions necessary to accrete objects in the primordial VZ would have
required extremely low eccentricities, or f$_{KE}$ significantly below 8\%, or both; low
eccentricities would in turn imply either the virtual absence of large perturbers inside 0.4 AU,
or the aid of some random velocity damping mechanism such as nebular gas-drag to prevent
self-stirring by a population of growing embryos, or both.  Whether the conditions necessary for
accretion in this region ever obtained is not clear, but the absence of a planet in this region
suggests that either accretion never proceeded very far in the VZ, or that any large (i.e.,
r$>$50 km) objects which formed were subsequently dynamically removed (a process which would
likely have contributed to further dynamical excitation and clearing of the region).

\medskip \noindent Despite our pessimism that accretion may have ever been able to proceed
in the VZ, for the remainder of this paper we posit that accumulation {\it did} take place
for a sufficient period some 4.5 Gyr ago to allow objects up to size scales with radii
near 50 km (the observational detection upper limit).  We then examine the constraints
that collisional models can place on the number of such objects that persist to the
present.

\centerline { }
\centerline { }
\centerline {\bf 3. POPULATION CONSTRAINTS FROM PRESENT-DAY} 
\centerline {\bf CATASTROPHIC DISRUPTION TIMESCALES}

\medskip \noindent A scale for the collisionality of the VZ region can be achieved from a
simple particle-in-a-box (PIB) estimate in which the collision time t$_{coll}$ on a target
of cross section $\sigma$ depends only on $\sigma$, the projectile number density $n$, and
the mean random speed of the projectiles $v$.  That is,

$$ 
t_{coll} = \left( n \sigma v \right)^{-1}.~~~\eqno (2) 
$$ 

\noindent For the assumption of just 10$^4$ radius r$>$0.1 km projectiles in a VZ extending from
0.09 AU to 0.21 AU with a mean random eccentricity of 10\%, we find that the mean time between
collisions on any given target with radius 3 km is $\sim$350 Myr; for a target with radius 30 km
the collision time is $\sim$3.5 Myr.  Although these are purely just {\it collision time}
estimates, given the erosive affects of such collisions (see \S2 above), one can conclude that
if there is a substantial population of objects larger than 3 km in the VZ, their fates will be
{\it strongly} affected by collisional evolution.

\medskip \noindent Although the simple collisionality estimate given above demonstrates the merit
of collisional considerations with respect to the Vulcanoid Zone, a far better first-order
assessment of the present-day collisional environment of the VZ can be derived using a static,
multi-zone collision {\it rate} model (CRM) which assesses the frequency of collisions in any
specified population.  This CRM code (Stern 1995; Durda \& Stern 2000) is based around a simple
but robust statistical PIB formalism, and computes orbit-averaged collision rates (and thus
collision timescales) for objects crossing heliocentric zones using accurate Kepler
time-of-flight calculations for the fractional time the target spends at each heliocentric zone
it crosses, depending on its orbital semi-major axis and eccentricity.  We used this model to
explore various plausible VZ population distributions as a function of heliocentric distance and
assumed mean random eccentricity (again, with $\langle i \rangle$=${1 \over 2}$$\langle e
\rangle$).  Our objective is to assess the range of collisional timescales onto targets which
observational constraints allow to exist in the VZ today.

\medskip \noindent Figure 3 depicts results of model runs for six plausible VZ cases spanning a
wide range of target strengths, $\langle e \rangle$'s, and VZ population.  In each of these cases
we assumed that the heliocentric surface mass density in the VZ declines like $R^{-2}$, where $R$
is heliocentric distance, and assumed the canonical -2.5 cumulative power law index Dohnanyi
collisional equilibrium population size distribution (e.g., Williams \& Wetherill 1994).  The
three CRM runs shown in Figure 3 assume populations of 10$^2$, 10$^3$, and 10$^4$ objects with
radius r$>$1 km in the VZ, respectively; the largest object in these three simulations (a direct
result of this population constraint and the Dohnanyi power law) is 4.0 km, 10.2 km, and 25.8 km
in radius, respectively.  Populations with significantly larger numbers of bodies with radii
larger than 1 km cannot exist in collisional equilibrium without violating observational
constraints.  The total mass of the three population ensembles, down to sizes of 0.1 km radius
was 1.9$\times$10$^{-9}$ M$_{\oplus}$, 2.3$\times$10$^{-8}$ M$_{\oplus}$, and
3.8$\times$10$^{-7}$ M$_{\oplus}$, respectively.\fnote {A set of 3 similar runs differing only in
that we assumed that the heliocentric surface mass density in the VZ declines like $R^{-1}$,
produced the same qualitative results.}

\medskip \noindent Consider now the catastrophic collisional disruption timescale results shown
in Figure 3.  The smallest projectile capable of disrupting and dispersing the largest object in
each given VZ population shown in Figure 3 is indicated along the collision timescale curves by
either a filled or open circle.  Filled circles are for the case of strong objects (both
projectiles and targets) in the VZ, and open circles are for the case of weak objects.  Here,
`strong' and `weak' are defined from the strongest and weakest of the published scaling laws in
the literature.  Specifically, the strongest scaling law is from Benz \& Asphaug (1999), and the
weakest is from Durda et al.~(1998).  The assumed specific disruption energies, $Q^*_D$, for
these cases are summarized in Table 1.

\vfill \eject

\centerline {\bf Table 1}
\centerline {\bf Scaling Law Specific Energies}

\def\hf{\hfil} 
\hskip 1. truein
\vbox{
\vskip\the\baselineskip 
\tabskip=0.5em
\halign{\hf #\hf & #\hf & #\hf \cr 
\noalign{\hrule} \cr 
\noalign{\hrule} \cr 

{\bf Target Radius} & {\bf $Q^*_D$ (strong)} & {\bf $Q^*_D$ (weak)} \cr 
\noalign{\hrule}\cr
\noalign{\hrule}\cr
04 km & 1$\times$10$^8$ ergs g$^{-1}$ & 1$\times$10$^6$ ergs g$^{-1}$ \cr
10 km & 2$\times$10$^8$ ergs g$^{-1}$ & 7$\times$10$^6$ ergs g$^{-1}$ \cr
25 km & 5$\times$10$^8$ ergs g$^{-1}$ & 6$\times$10$^7$ ergs g$^{-1}$ \cr
\noalign{\hrule} \cr 
\noalign{\hrule} \cr}}

\medskip \noindent Even in the smallest of these three VZ population scenarios (upper
panel), which has just 100 objects with radii larger than 1 km in the entire VZ, the
catastrophic collisional disruption timescale of the largest object in the swarm, a 4 km
radius body, is less than the age of the solar system; this result obtains over the full
range of $\langle e \rangle$ explored, throughout the VZ if the target is mechanically weak.
If the target is mechanically strong, this result obtains out to a heliocentric distance of
0.17 AU. Strong objects at larger heliocentric distance survive longer owing to a
combination of lower collision rates and lower collision velocities (thus requiring
progressively larger impactors to cause disruption); the latter factor dominates this
progression.  Since collision timescales increase with target radius squared in this size
regime, smaller objects have collisional disruption timescales that are longer than those
shown here for the 4 km target by the ratio (4 km/r)$^2$, implying that there is a
significant region of strength-heliocentric distance parameter space for objects of 0.1 km
to 1 km scale to survive against both collisions and PR drag.\fnote {Recall that a 0.1 km
objects of density 4 g cm$^{-3}$ will spiral from the outer limit of the VZ at 0.21 AU to
its evaporation limit near 0.07 AU in 4.5 Gyr.}  Qualitatively similar behaviors are seen
for the two larger hypothetical VZ population runs, which have their results depicted in the
lower two panels of Figure 3, respectively.

\medskip \noindent From Figure 3 we conclude that few if any objects with radii of $\approx$1--25
kilometers are likely to survive against collisions for the age of the solar system in standard
population structures like the ones we explored, a result in accord with the observational
absence of objects in the 10 km to 50 km size range.  This result is not unexpected, of course,
because the volume of space in the VZ is so small and the orbit speeds are so high.

\medskip \noindent From the results shown in Figure 3, one concludes that virtually no {\it
primordial} objects with 1 km$<$r$<$25 km could have survived to the present in the low-mass VZ
models we have considered here.  Higher mass models with the same population structure would be
even more collisional.

\medskip \noindent One could imagine scenarios, however, in which larger objects formed or were
transported into the region, and then subsequently suffered collisional erosion owing to the
growth of orbital eccentricities.  To model such populations it is necessary to use
time-dependent collisional evolution simulations.  We discuss such simulations next.

\centerline { }
\centerline { }
\centerline {\bf 4. TIME-DEPENDENT COLLISIONAL EVOLUTION SIMULATIONS}

\medskip \noindent The time-dependent model we use to investigate collisional evolution in
the Vulcanoid Zone was adapted from the Kuiper Belt code described by Stern \& Colwell
(1997), with its target mechanical properties changed to reflect the range of likely object
types in the Vulcanoid Zone.  Very briefly, this model uses a ``moving bin'' (i.e.,
Lagrangian) approach to size bins first described by Wetherill (1990). This technique has the
advantage of being particularly straightforward, and particularly accurate in its mass accounting.  The
radius bins we used were separated by factors of 2$^{1/3}$; we ran bin radii from 1 m to
100 km for the initial VZ simulations.  Three-body, Keplerian shear-limited, gravitational
collision cross sections were computed following Ward's (1996) prescription.  For each
collision pair of mass $m_k$$<$$m_l$ colliding at relative velocity $v_{kl}$, the specific
impact energy is computed according to the standard definition, Q$^*$=${{1} \over
{2}}m_kv_{kl}^2/m_l$ (e.g., Housen \& Holsapple 1990), and then compared to a threshold
value for catastrophic disruption, $Q^*_D$.  We used a strain-rate scaling
model (Housen \& Holsapple 1990; HH90); this results in objects slightly stronger than
the strong cases in \S3, above.

\medskip \noindent For all impacts, we initially add the mass of the impactor to the target, and
then remove the appropriate amount of debris, based on the target and impactor properties and the
collision energetics.  The result is net accretion if the mass of the escaping ejecta is less
than the impactor mass, and net erosion if the ejected mass exceeds the mass of the impactor.
Our catastrophic fragmentation model is the same as that used in Colwell \& Esposito (1993).  If
Q$^*$$>$Q$^*_D$, then the mass fraction with escape velocity from the colliding pair is given by
$f$($>v_{\rm esc}$)=$1/2(v_{\rm esc}/v_{\rm med})^{-3/2}$, where $v_{med}$=$\sqrt{2f_{KE}Q^*}$ is
the median fragment velocity, $f_{KE}$ is the fraction of impact energy partitioned into fragment
kinetic energy, and $v_{\rm esc}$ is the escape velocity.  Following experimental results (see
Fujiwara et al.~1989), we have set $f_{KE}$=0.10.\fnote {Since, particularly for larger objects,
lower values of $f_{KE}$ might be more appropriate, we note that any significant lowering of
$f_{KE}$ would lengthen collisional erosion and catastrophic disruption timescales.  We have
found that this lengthening scales somewhere between 1/$f_{KE}$ and 1/$\sqrt{f_{KE}}$, depending
on target size and bulk mechanical properties.}

\medskip \noindent In our model the total mass of escaping debris is distributed to smaller mass
bins following a standard, two-component power-law size distribution, with slopes computed based
on laboratory experiments (e.g., Davis \& Ryan 1990).  The result of any given collision can
range from complete accretion (no debris achieves escape velocity from the colliding pair), to
complete erosion (in which the object is destroyed because greater than half the target mass has
escape velocity).  Cratering impacts ($Q^*$$<$$Q^*_D$) are handled similarly:  following the
literature, the debris size distribution is a single-valued power-law
(n($>$m)$\propto$$m^{-5/6}$); the fragment velocity distribution uses a power law exponent of
$-1.2$ (for weak target runs) and $-2.0$ (for our hard target runs).  Debris smaller than the
smallest discrete bin (1 meter in the initial runs presented below) is placed into a ``dust''
bin.  PR drag operates to remove small debris from the simulation, based on their size- and
$a$,$e$-dependent PR drag lifetime.  Runs of the model typically conserve mass to a few parts in
10$^{-15}$ over 10$^{10}$ years.

\medskip \noindent Figure 4 presents the first of several sets of VZ collisional evolution
simulations we performed using this model.  For all of the runs presented below we assumed
uniform density and impact strengths of 4 gm cm$^{-3}$ and 3$\times$10$^6$ ergs gm$^{-1}$,
respectively, for all objects in the simulation.  This combination of density and strength
corresponds to competent basalt, i.e., somewhat stronger objects than in HH90.  Figure 4 presents
various results for objects with semi-major axis a=0.20 AU.  Each simulation was evolved until it
achieved an end state with no objects larger than 1 m among the population with a=0.20 AU.  Note,
however, that the plots and timescales in Figure 4 refer to the time at which the given
population evolves to have no objects with r$>$1 km; the loss of smaller objects proceeds rapidly
after this point.  The purpose of these collisional evolution simulations was to explore the
evolution of various populations that fit with the available VZ observational constraints
reviewed in \S1.

\medskip \noindent The set of simulations shown in the upper two panels of Figure 4 were started
with 300 objects with r$>$1 km in the VZ (i.e., between 0.06 and 0.21 AU); smaller debris down to
r=1 m in radius was extrapolated from the large object population using a Dohnanyi cumulative
power-law population index of -2.5.  The simulations shown in the lower two panels started with
10$^4$ objects with r$>$1 km in the VZ.  The two left-hand panels in Figure 4 refer to an
assumed, constant mean random eccentricity $\langle e \rangle$=0.0032; the two right-hand panels
assume $\langle e \rangle$=0.1024.  Owing to the high Kepler velocities at 0.20 AU, even for the
strong objects assumed here, both population cases are erosive across all of the populated size
bins.

\medskip \noindent As stated above, the two cases shown in the upper panels were started with
only 300 objects in the entire VZ larger than 1 km radius; this corresponds to a starting
condition with just 8 objects larger than 1 km radius (largest object r=2.05 km) in our 0.01 AU
wide bin centered at a=0.20 AU.  While the $\langle e \rangle$=0.1024 case took only 1.2 Gyr yrs
to eliminate all objects with r$>$1 km, the $\langle e \rangle$=0.0032 case took 6.3 Gyr.
Examining the intermediate-time population structures in each of these cases reveals differing
population structure erosion styles.  In the case with $\langle e \rangle$=0.1024, high energy
collisions by the numerous small bodies quickly destroyed the largest objects (i.e., r$>$100 m)
through catastrophic collisions.  The case with $\langle e \rangle$=0.0032, though still erosive,
was not sufficiently energetic to induce rapid catastrophic impacts on the largest objects, and
resulted in a more gradual erosion of the population structure throughout the run.

\medskip \noindent Now consider the two cases shown in the lower panels of Figure 4, i.e., the
runs with 10$^4$ objects in the VZ larger than 1 km radius at the simulation start.  This
corresponds to a starting condition with 272 objects larger than 1 km radius (largest object
r=8.2 km) in our 0.01 AU wide bin centered at a=0.20 AU.  In the case with $\langle e
\rangle$=0.1024, the system population number density was so high and the collisions so energetic
that in just 1.6 Myr it evolved to a state with no objects with radius larger than 1 km.  In the
case with $\langle e \rangle$=0.0032, however, this evolution did not obtain in the 10 Gyr length
of the simulation.  However, after 4.5 Gyr this run contained only 9 objects with r$>$1 km.  As
shown in Figure 5, additional runs with both 300 and 10$^4$ objects initially in the VZ with
r$>$1 km, but started with intermediate $\langle e \rangle$'s of 0.012 to 0.025, yielded
timescales of 1 Gyr to 3 Gyr to erode down to populations with no objects with r$>$1 km and
a=0.20$\pm$0.005 AU.

\medskip \noindent A suite of runs just like those in the upper two panels of Figure 4 but with
$\langle e \rangle$=0.0004 also produced erosion.  Though the timescale to fully deplete the
population of 1 km objects at 0.20 AU exceeded the age of the solar system, only 14 objects with
r$>$1 km remained from a starting Dohnanyi population with 10$^4$ r$>$1 km objects after 4.5 Gyr.
Figure 5 presents a similar set of run results for cases with $\langle e \rangle$=0.0124 and
$\langle e \rangle$=0.0256, respectively.

\medskip \noindent Together these various results at 0.20 AU indicate that, except in the case
where $\langle e \rangle$ can be maintained significantly below 4$\times$10$^{-4}$ (i.e., below
the $e^*$ boundary), any substantial VZ population near 0.20 AU extending up to objects with
radii of a few tens of km must be collisionally eroded by the present 4.5 Gyr age of the solar
system to a point where r=1 km and larger objects are either rare or non-existent.\fnote {We also
conducted a set of simulations identical to those in Figure 4, but removed all objects with r$<$1
km from the starting populations.  Though evolution proceeded more slowly at first owing to the
need to build up the population of small projectiles from collisions among km-sized and larger
bodies, in all 4 cases run, we again found that the population of objects with r$>$1 km was
reduced to 10 or less objects remaining at a=0.20 AU over the age of the solar system.}

\medskip \noindent One of course expects more rapid evolution at smaller heliocentric distance,
owing to number density enhancements and increased collision energetics.  Figure 6 shows a set of
simulations identical to those in Figure 4, but at a=0.10 AU.  The resulting evolutions in
population size structure are qualitatively similar, but with the timescales accelerated by
factors of 7 to 10.

\medskip \noindent Weaker mechanical properties, steeper initial power-law population ensembles,
higher mean random eccentricities, higher inclinations with respect to eccentricity, and the
inclusion of a bombarding flux from the asteroid belt and cometary reservoirs would each shorten
the VZ erosion timescales quoted above.\fnote {Regarding collisions with objects on heliocentric
orbits outside the VZ, we find that such collisions are rare, and that collisional lifetimes
exceed the age of the solar system.  More specifically, based on Levison et al.'s (2000) cometary
impact rates on Mercury, we find that the catastrophic collision lifetime for objects down to 1
km in radius is in excess of 10$^{10}$ years in the center of the VZ; for 10 km radius targets
in the center of the VZ the estimated catastrophic collision lifetime exceeds 4$\times$10$^{11}$
years.}  The Yarkovsky effect increases the rate of small debris transport and removal over PR
drag alone (e.g., Farinella et al.~1998, Vokrouhlick\'y 1999), thereby reducing the impact flux
on larger objects in the VZ.  Therefore, this effect, though not modelled here owing to its wide
range of free parameter choices, will tend to moderately increase the estimated lifetimes of
bodies with diameters of several km and larger, but may actually allow objects as large as 1 km
in diameter to be dragged into the Sun, even from 0.2 AU over 4.5 Gyr (W.~Bottke,
pers.~comm.~1999).

\medskip \noindent Our results demonstrate that unless f$_{KE}$ is substantially below 10\%, or
the VZ has been maintained below the $e^*$ boundary, i.e., $\langle e \rangle$ below
$\sim$10$^{-4.5}$, population structures that fit under the present-day observational constraint
boundaries would self destruct owing to collisions, resulting in a VZ which today is so thinly
populated that collisions are rare.  This population constraint implies that a few tens, and
quite likely a very much smaller number of objects in the 1 km to 50 km size range are extant
today inside 0.20 AU.

\medskip \noindent Figure 7 presents some results from one additional set of simulations we
performed at 0.20 AU.  In this set of cases we examined the evolution of much more massive
starting populations with objects as large as r=330 km.  Although observations trivially rule out
such a massive VZ population today, it is instructive to examine what the evolution of such
swarms would be, so as to determine what if any signatures of such a primordial population might
exist today.  In these simulations the total mass of the VZ down to our cutoff radius (10 m in
this case) was 9$\times$10$^{-4}$M$_{\oplus}$, i.e., somewhat in excess of the present-day mass
of the asteroid belt (7$\times$10$^{-4}$M$_{\oplus}$).

\medskip \noindent Specifically, the lefthand panel in Figure 7 shows the gentlest and slowest
evolving of the four runs we performed in this scenario.  In this case we assumed $\langle e
\rangle$=0.0032; this is a low enough eccentricity to actually allow the largest object in the
starting population to grow.  The end result of this run was that after 22 Myr, two thirds of the
starting mass in this zone had been removed from the simulation owing to grinding and subsequent
PR drag loss.  Further, after 22 Myr, no objects remained in our standard 0.01 AU wide model zone
at 0.20 AU with r$>$1 km, except a single, largest body that had grown to r=371 km.  After 45
Myr, no objects remained with r$>$0.1 km, except the large object which was stranded in the
population but which could not grow appreciably because there was so little mass left in the
population of small debris.  Of course, this simulation is not fully self-consistent in that we
simply began with large bodies up to 330 km in radius at the start.  As the results in Figure 2
show, achieving growth from km-scale and smaller bodies to this stage, requires maintaining
$\langle e \rangle$'s an order of magnitude or more lower than in this run until objects with
r$\sim$10 km are grown.

\medskip \noindent Runs starting with the large bodies up to r=330 km but with higher $\langle e
\rangle$ produced much greater quantities of debris owing to the more energetic collisions, and
therefore evolved much faster.  For example, the righthand panel in Figure 7 shows a run with
$\langle e \rangle$=0.0128.  In 18 Myr this ensemble ground away 80\% of its mass, and contained
only 33 objects with r$>$1 km, the largest of which had r=34 km.  As noted above, these runs were
performed at 0.20 AU heliocentric distance; we found that evolution proceeds about an order of
magnitude faster still at 0.10 AU heliocentric distance.

\medskip \noindent The more massive VZ scenarios just described demonstrate that even if the VZ
was able to create a Vulcanoid belt of similar scale to the asteroid belt early in the history of
the solar system, it would by today either have been eroded away (if $\langle e \rangle$ exceeded
$e^*$ for as little as 1\% the age of the solar system), or (if $\langle e \rangle$ remained well
below $e^*$) it would have grown a small number of larger objects which are not seen today.  Had
that latter condition occurred, dynamical stability results (e.g., Evans \& Tabachnik 1999) imply
that one of more of these objects would remain and have been detected.

\centerline { }
\centerline { }
\centerline {\bf 5. CONCLUSIONS}

\medskip \noindent We have examined the role of collisional evolution in the Vulcanoid Zone (VZ),
where searches for a population of small bodies have been conducted several times.  The Vulcanoid
Zone, owing to its shorter dynamical times and smaller volume is far ``older'' collisionally (and
dynamically) than the asteroid belt.  Unless our $f_{KE}$=0.1 is a gross overestimate, or the
Vulcanoids are far denser or stronger than our adopted values, then:

\medskip \item {$\bullet$} If the mean random orbital eccentricity exceeds a critical value,
$e^*$ (a function of target mass, mechanical properties, and heliocentric distance), efficient
collisional grinding and erosion must take place.  Given that the largest objects which
observations allow to exist in the VZ today has a radius near 30 km, this implies that $e^*$ is
today less than a few times 10$^{-3}$, and could be an order of magnitude smaller if the largest
bodies in the VZ are only a few km in radius.

\medskip \item {$\bullet$} Collisional grinding and the subsequent radiation transport of debris
out of the VZ dramatically depletes starting populations that are consistent with the existing
observational constraints (i.e., VZ masses $\sim$10$^{-6}$M$_{\oplus}$, largest objects with
r$\approx$25 km).  This obtains whether one starts the evolution with or without a Dohnanyi-like
debris tail of objects.  This evolution results in populations which, unless eccentricities are
below $\sim$10$^{-3}$, cannot contain more than a few hundred objects with radii exceeding 1 km.

\medskip \item {$\bullet$} Even allowing for ancient VZ ensembles with collisional equilibrium
power-law population structures and embedded objects up to 330 km in radius (i.e., leading to a
mass somewhat in excess of the asteroid belt), collisional evolution is so fast and collision
energies are so high, that populations with mean random orbital eccentricities above
$\sim$3$\times$10$^{-3}$ will ``self-destruct'' down to levels with only a residuum of widely
spaced (and therefore collisionally non-interacting bodies) in $\sim$1\% the age of the solar
system.  In our simulations, this residuum contained only a few hundred objects across the entire
VZ with r$<$1 km for orbits with $\langle e \rangle$ near 10$^{-2}$.

\medskip \item {$\bullet$} Collisional evolution will proceed most quickly at smaller
heliocentric distances; this, combined with PR drag and the Yarkovsky effect will cause any
former or present-day Vulcanoid Zone population to be depleted by collisional grinding from the
``inside out'' over time.

\medskip \item {$\bullet$} The characteristic erosion timescale for the VZ can range from 10$^7$
yrs to 10$^{10}$ years, depending on $\langle e \rangle$, $f_{KE}$, and the initial population
density.  Therefore, a wide range of VZ erosion timescales may exhibit themselves in solar
systems with architectures like our own.  The observational signatures of VZ erosion, i.e.,
thermal emission at $\sim$2--5 $\mu$m and photospheric pollution with silicate-iron signatures
may someday be detected in other planetary systems.

\medskip \item {$\bullet$} These considerations suggest it is unlikely that, unless we have
grossly overestimate $f_{KE}$, more than a few hundred objects with radii larger than 1 km will
be found in the VZ.  The most favorable location to search for such bodies is in highly circular
orbits near the outer edge of the dynamically stable VZ (i.e., near 0.2 AU), where collisional
evolution and radiation transport of small bodies and debris proceed most slowly.

\medskip \noindent Although our exploration of parameter space is not fully complete (e.g., we
did not examine scenarios with 1000 km radius and larger bodies in the starting population), we
do believe that the work discussed here shows that the present-day VZ is likely to be either
depleted or almost depleted of km-scale and larger objects.  If any such objects are found, then
collisional evolution arguments imply it is highly likely that their number density will be so
low, and their spacings so great, that they will form a thin, collisionally-decoupled population
remnant from an ancient era.

\medskip \noindent In conclusion, our work suggests that large numbers of objects with radii of
km scale or larger are unlikely to be found unless the VZ region of the solar system has never
been dynamically excited to orbital eccentricities above $\sim$10$^{-3}$, which seems unlikely.
Nevertheless, the detection of {\it any} such population, regardless of how low, would shed valuable
light on the dynamical, and possibly the accretional/erosional, history of this end-member region
of our solar system, and would no doubt bear on our understanding of extra-solar planetary
systems as well.

\vfill \eject

\centerline {\bf ACKNOWLEDGMENTS}

\medskip \noindent This research was supported by the NASA Origins of Solar Systems Program and
the NASA Sun-Earth Connection Guest Investigator Program.  We thank Bill Ward for several
interesting conversations.  We thank Luke Dones, Michel Festou, and Joel Parker for useful
comments on our draft manuscript, and Clark Chapman, both for a careful reading of the manuscript
and for useful suggestions concerning the selection of model runs.  We also thank our referees at
Icarus, Erik Asphaug and David Rubincam, for their helpful feedback.

\vfill
\eject

\centerline {\bf REFERENCES}

\medskip \noindent Benz, W., and E.~Asphaug, 1999.  Catastrophic disruptions revisited.
{\it Icarus}, in press.

\medskip \noindent Campins, H.H., D.R.~Davis, S.J.~Weidenshilling, and M.~Magee, 1996.
Searching for Vulcanoids.  In {\it Completing the Inventory of the Solar System}. ASP
Confr.~Series, Vol.~{\bf 107} (T.~Rettig \& J.~Hahn, eds.), 85--96.

\medskip \noindent Colwell, J.E., and L.W.~Esposito, 1993.  Origin of the rings of Neptune
and Uranus II.  Initial conditions and ring moon populations.  {\it JGR}, {\bf 98},
7387--7401.

\medskip \noindent Campbell, W.W., and R.J.~Trumpler, 1923.  Search for intramercurial bodies.
{\it PASP}, {\bf 35}, 214--216.

\medskip \noindent Courten, H., 1976.  Ten years of solar eclipse comet searches.  {\it BAAS},
{\bf 8}, 504.

\medskip \noindent Davis, D.R., and E.V.~Ryan, 1990.  On collisional disruption:
Experimental results and scaling laws.  {\it Icarus}, {\bf 83}, 156--182.

\medskip \noindent Durda, D.D., and S.A.~Stern, 2000.  Collision rates in the present-day
Kuiper Belt and Centaur regions:  Applications to surface activation and modification on
comets, KBOs, and Pluto-Charon.  {\it Icarus}, submitted.

\medskip \noindent Durda, D.D., R.~Greenberg, \& R.~Jedicke, 1998.  Collisional models and
scaling laws:  A new interpretation of the shape of the main-belt asteroid size distribution.
{\it Icarus}, {\bf 135}, 431--440.

\medskip \noindent Evans, N.W., and S.~Tabachnik, 1999.  Possible long-lived asteroid
belts in the inner solar system.  {\it Nature}, {\bf 399}, 41--43.

\medskip \noindent Farinella, P., D.~Vokrouhlick\'y, and W.K.~Hartmann, 1998.  Meteorite
delivery via Yarkovsky orbital drift.  {\it Icarus}, {\bf 132}, 378--387.

\medskip \noindent Fujiwara, A., et al., 1989.  Experiments and scaling laws in catastrophic
collisions.  {\it Asteroids II} (R.P.~Binzel, T.~Gehrels, and M.S.~Matthews, eds.).
U.~Az.~Press, Tucson, 240--268.

\medskip \noindent Hodapp, K.-W., R.M.~MacQueen, and D.N.B.~Hall, 1992.  A search during the
1991 solar eclipse for the infrared signature of circumsolar dust.  {\it Nature}, {\bf 355},
707--710.

\medskip \noindent Housen, K.R., and K.A.~Holsapple, 1990.  On the fragmentation of asteroids and
planetary satellites.  {\it Icarus} {\bf 84}, 226--253.

\medskip \noindent Leake, M., C.R.~Chapman, S.J.~Weidenshilling, D.R.~Davis, and
R.~Greenberg, 1987.  The chronology of Mercury's geological and geophysical evolution:
The Vulcanoid hypothesis.  {\it Icarus}, {\bf 71}, 359--375.

\medskip \noindent Lebofsky, L.A., 1975.  The stability of frosts in the solar system.  {\it
Icarus}, {\bf 25}, 205--217.

\medskip \noindent Leinert, C., I.~Richter, E.~Pitz, and B.~Planck, 1981.  The zodiacal light
from 0.1 to 0.3 AU as observed by the Helios space probes.  {\it A\&A}, {\bf 103}, 177--185.

\medskip \noindent Levison, H.F., M.J.~Duncan, K.~Zahnle, \& L.~Dones, 2000.  Planetary
impact rates from ecliptic comets.  {\it Icarus}, in press.

\medskip \noindent MacQueen, R.M., and B.W.~Greely, 1995.  Solar coronal dust scattering in
the infrared. {\it ApJ}, {\bf 440}, 361--369.

\medskip \noindent Perrine, C.D., 1902.  Results of the search for an intra-Mercurial
planet in the total solar eclipse of 1901, May 18.  {\it Lick.~Observe.~Bull.}, {\bf 1},
183--187.

\medskip \noindent Rettig, T.W., and J.M.~Hahn, 1996.  Completing the Inventory of the Solar
System.  {\it ASP Conference Series}, {\bf 107}, San Francisco, 395pp.

\medskip \noindent Stern, S.A., 1995.  Collision rates in the Kuiper disk and their
implications.  {\it AJ}, {\bf 110}, 856--865.

\medskip \noindent Stern, S.A., and J.E.~Colwell, 1997. Collisional erosion in the 
Edgeworth-Kuiper belt. {\it ApJ}, {\bf 490}, 879--884.

\medskip \noindent Veverka, J., P.~Helfenstein, B.~Hapke, and J.D.~Goguen, 1988.  Photometry
and polarimetry of Mercury.  In {\it Mercury} (F.~Vilas, C.R.~Chapman, and M.S.~Matthews,
eds.).  U.~Az.~Press, Tucson, 37--58.

\medskip \noindent Vokrouhlick\'y, D.  1999.  A complete linear model for the Yarkovsky thermal
force on spherical fragments.  {\it A\&A}, 433, 362--266.

\medskip \noindent Ward, W.R., G.~Colombo, and F.~Franklin, 1976.  Secular resonance, solar
spin down, and the orbit of Mercury.  {\it Icarus}, {\bf 28}, 441--452.

\medskip \noindent Ward, W.R., 1996.  Planetary accretion.  In {\it Completing the
Inventory of the solar system} (T.~Rettig and J.~Hahn, eds.), ASP Conference Series, {\bf
107}, 337--360.

\medskip \noindent Wetherill, G.W., 1990.  Comparison of analytical and physical modeling of
planetesimal accumulation.  {\it Icarus}, {\bf 88}, 336--354.

\medskip \noindent William, D.R., and G.W.~Wetherill, 1994.  Size distribution of
collisionally evolved asteroid populations:  Analytical solution for cascades.  {\it
Icarus}, {\bf 107}, 117--128.

\vfill
\eject

\centerline {\bf FIGURE CAPTIONS}

\medskip \noindent {\it Figure 1.}  Observational constraints on possible Vulcanoid sizes
as a function of heliocentric distance and assumed surface geometric albedo.  For the
magnitude-limited searches, radii have been calculated assuming a visual magnitude V=8.5,
observations at quadrature (phase angle 90 deg), a phase function like that of Mercury
(Veverka et al.~1988), and three different albedos:  p=0.05 (dark asteroidal), p=0.14
(Mercury), and p=0.30 (bright asteroidal).

\medskip \noindent {\it Figure 2.}  Critical collision eccentricities separating the
erosive vs.~accumulation regimes (see \S 2) are computed here as a function of target
radius for both strong and weak target mechanical properties at 0.06 and 0.21 AU in the
Vulcanoid Zone.  Strong target parameters:  density $\rho$=4 gm cm$^{-3}$ and the material
shattering strength $Q^*_S$=3$\times$10$^6$ ergs gm$^{-1}$; weak target parameters:
density $\rho$=1 gm cm$^{-3}$ and material shattering strength $Q^*_S$=3$\times$10$^4$
ergs gm$^{-1}$.  Note:  For bodies with non-negligible binding energy (i.e., mass), the
shattering strength Q$^*_S$ is less than the disruption strength Q$^*_D$ because to
disrupt such a body also requires removing half its mass to infinity.

\medskip \noindent {\it Figure 3.}  Collisional timescale results, as a function of
projectile radius and heliocentric distance, using the static, multi-zone collision rate
model, for the three population cases described in \S3 of the text.  The horizontal dashed
line is a timescale of 4.5 Gyr.  The sloping collision timescale lines are shown at R=0.09
AU and a=0.21 AU.  The dotted line cases assume $\langle e \rangle$=0.2048, and the solid
lines assume $\langle e \rangle$=0.0256; together these two cases span a wide range of
potential VZ eccentricities.  The circles on each of these collision rate curves represent
the boundary between cratering and catastrophic collisions; open circles correspond to the
assumption of weak targets and filled circles correspond to the assumption of strong
targets (see text).

\medskip \noindent {\it Figure 4.}  Vulcanoid Zone collisional evolution simulations for
objects with a=0.20 AU.  The simulations shown in the upper two panels began with 300
objects with r$>$1 km in the entire 0.06--0.21 AU VZ.  The simulations shown in the lower
two panels began with 10$^4$ objects with r$>$1 km in the entire 0.06--0.21 AU VZ.  The
two cases on the left assume $\langle e \rangle$=0.0032; the two cases on the right assume
$\langle e \rangle$=0.1024.  The dotted line is the initial population.  The successively
thicker, solid lines represent the population at 3\%, 10\%, 30\%, and 100\% of the run
time shown, where the run time is the simulation time required to reach a state with no
objects with radius $>$1 km in the 0.25 AU subzone.  See text for additional simulation
details.

\medskip \noindent {\it Figure 5.}  Same as Figure 4, but for $\langle e \rangle$=0.0124
and $\langle e \rangle$=0.0256.

\medskip \noindent {\it Figure 6.}  Same as Figure 4, but for a=0.10 AU.

\medskip \noindent {\it Figure 7.}  Collisional evolution runs at 0.20 AU for massive VZ
scenarios (total mass 9$\times$10$^{-4}$M$_{\oplus}$).  At the simulation start, the
population (as shown by the dotted line) contains objects up to 330 km in radius, and as
small as 20 m in radius, connected by a Dohnanyi power-law ensemble of intermediate objects.
Left panel:  $\langle e \rangle$=0.0032; right panel:  $\langle e \rangle$=0.0128.  See text
for discussion.

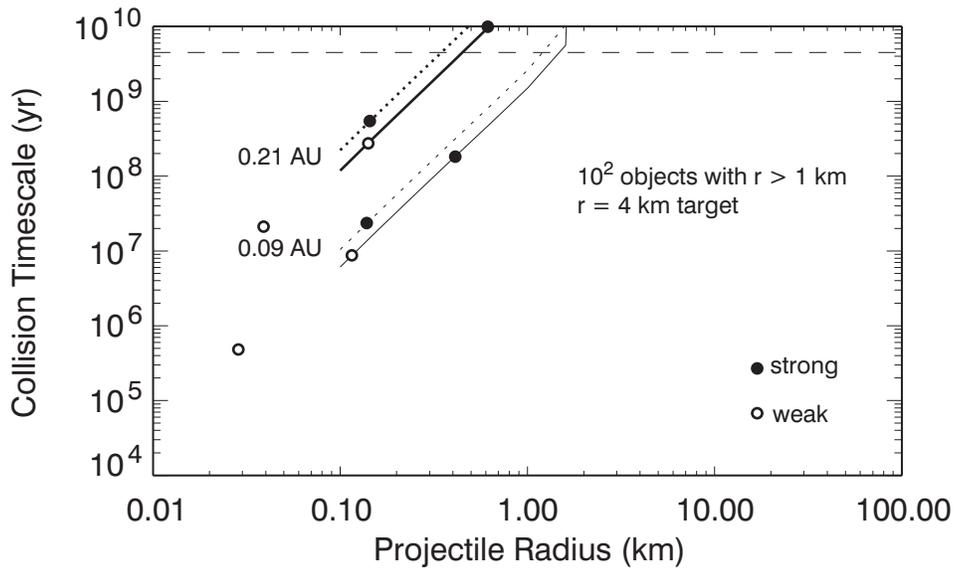
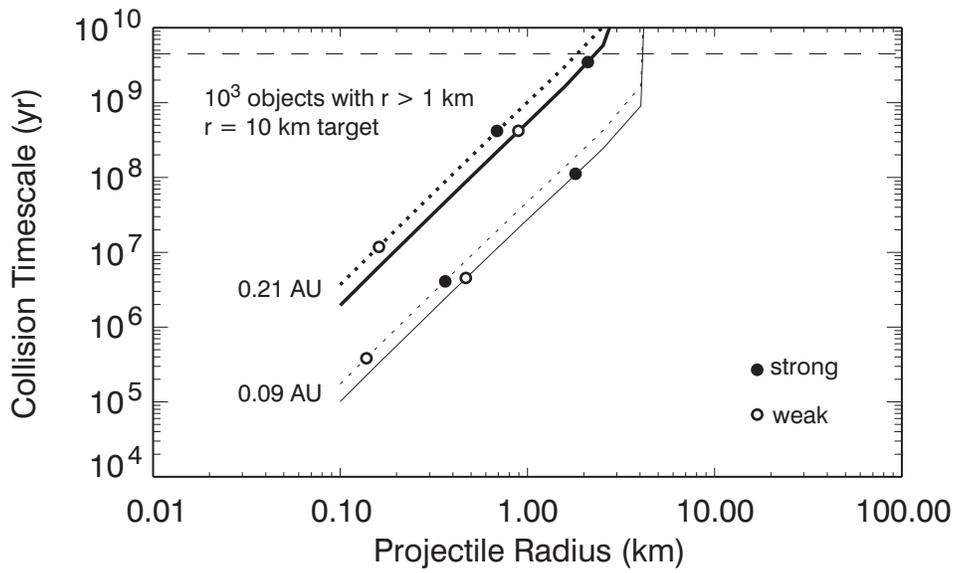
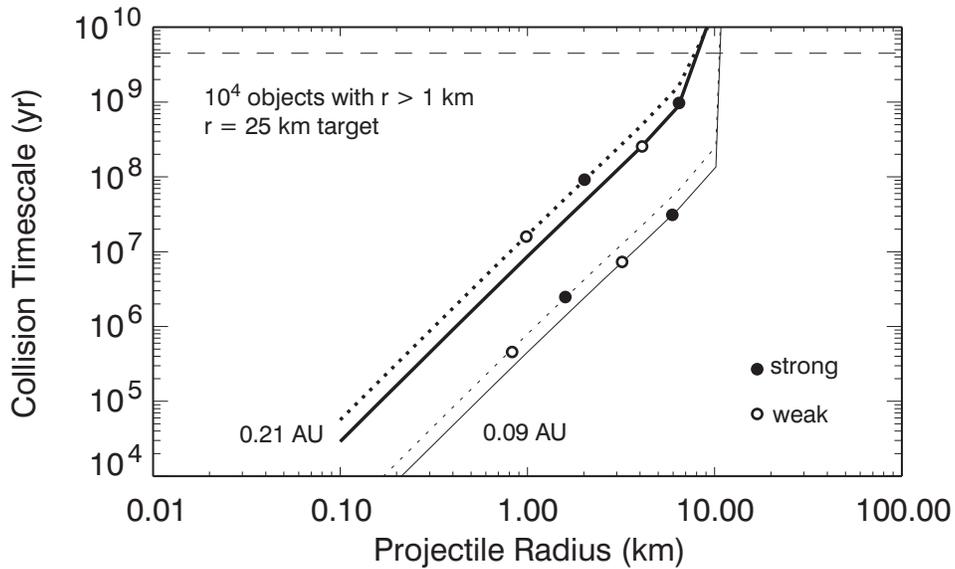
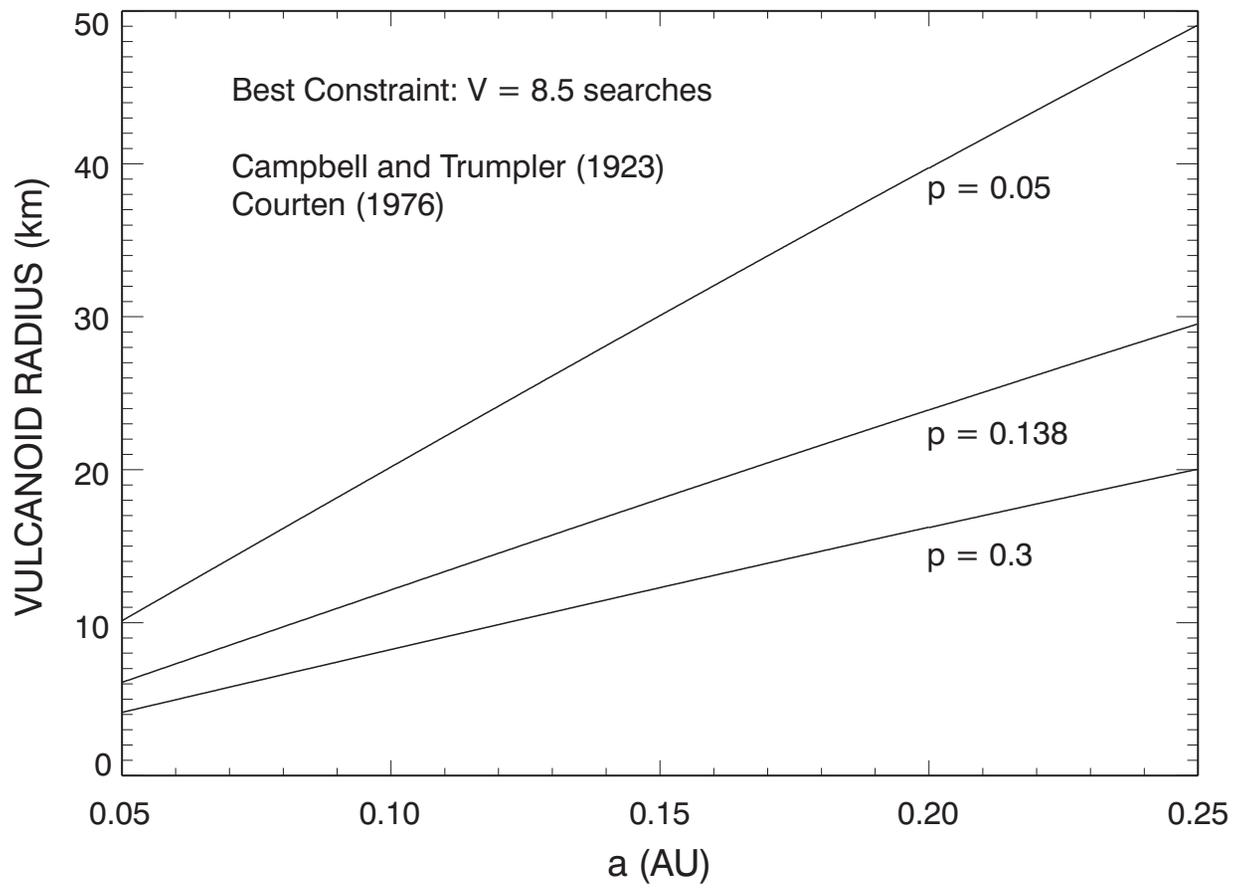
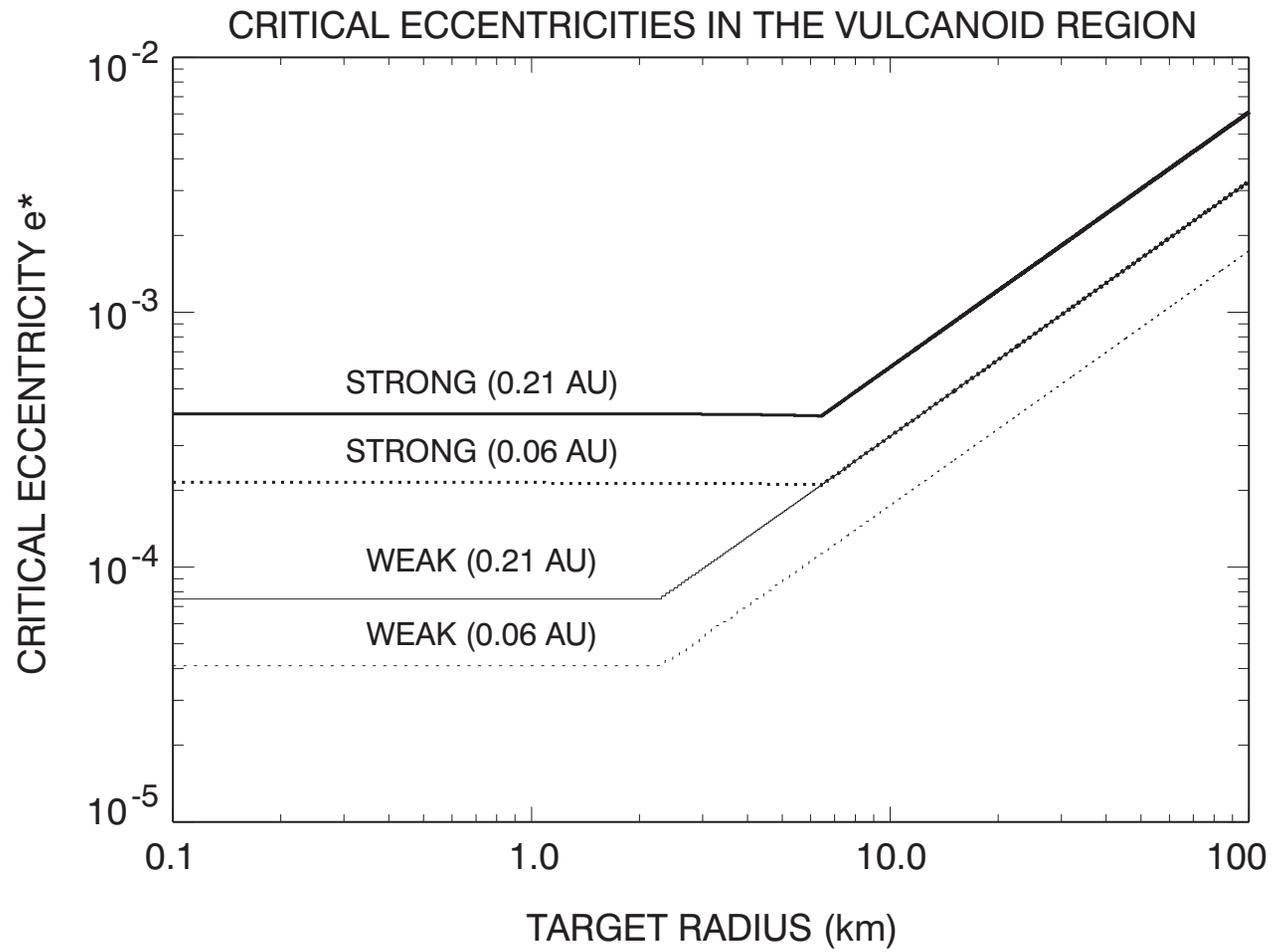
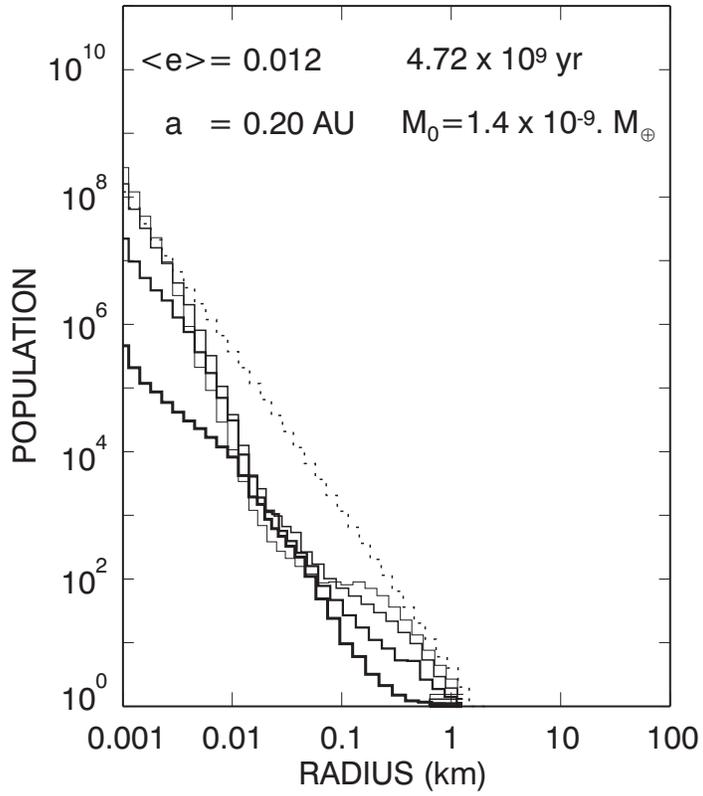
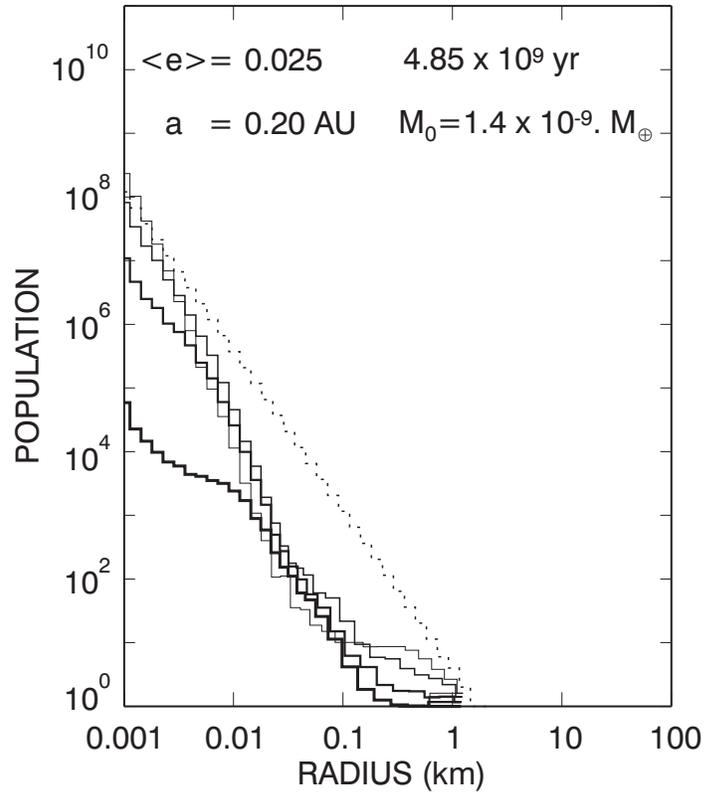
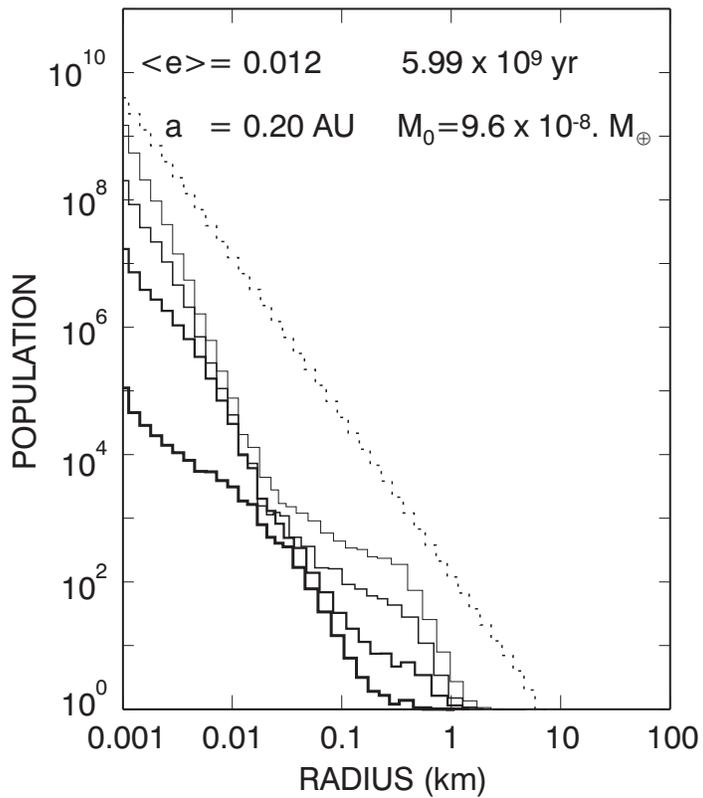
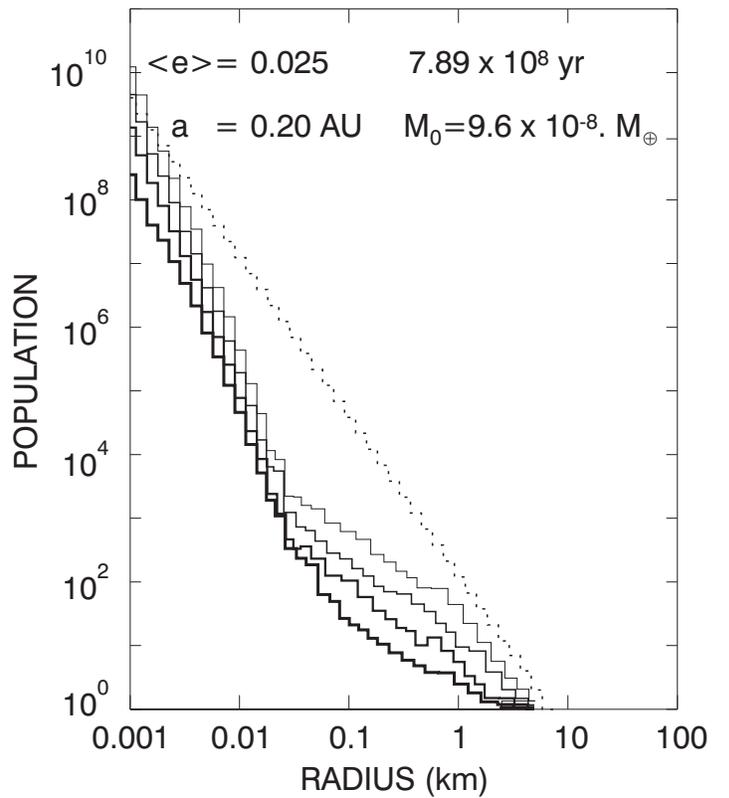
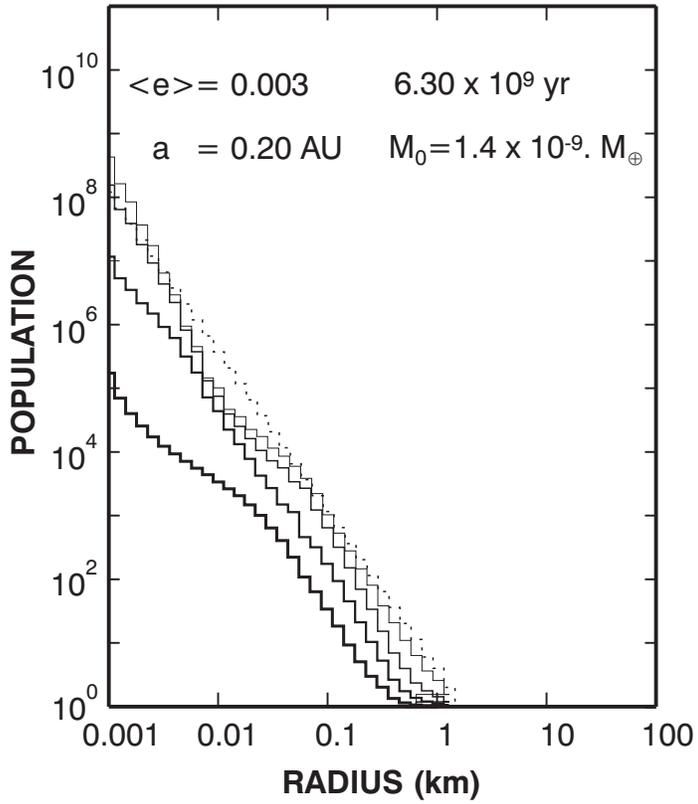
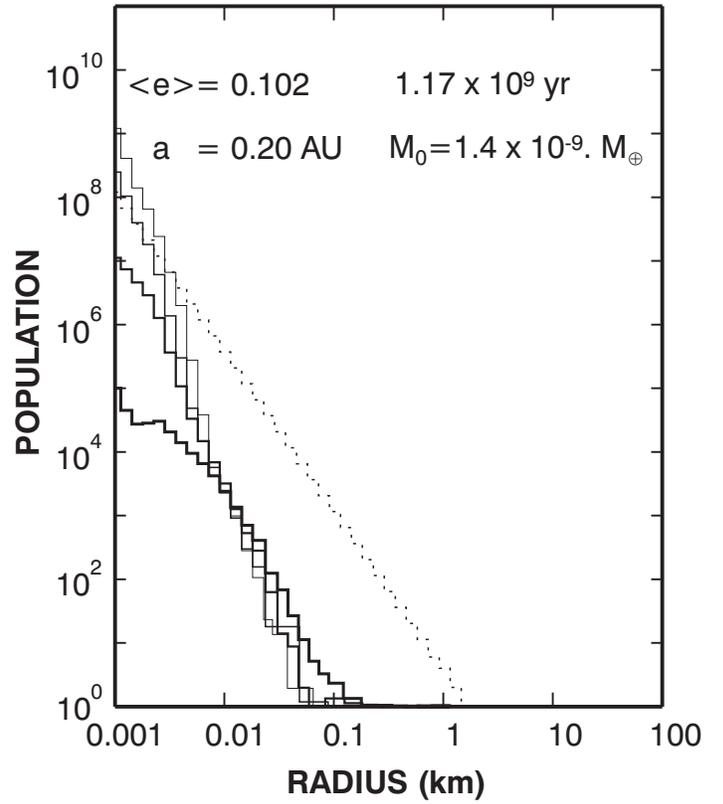
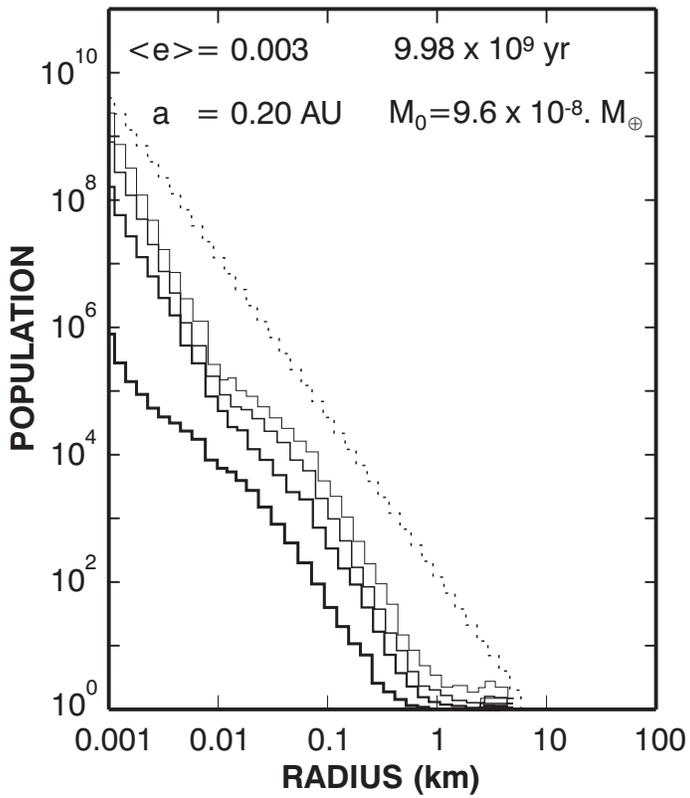
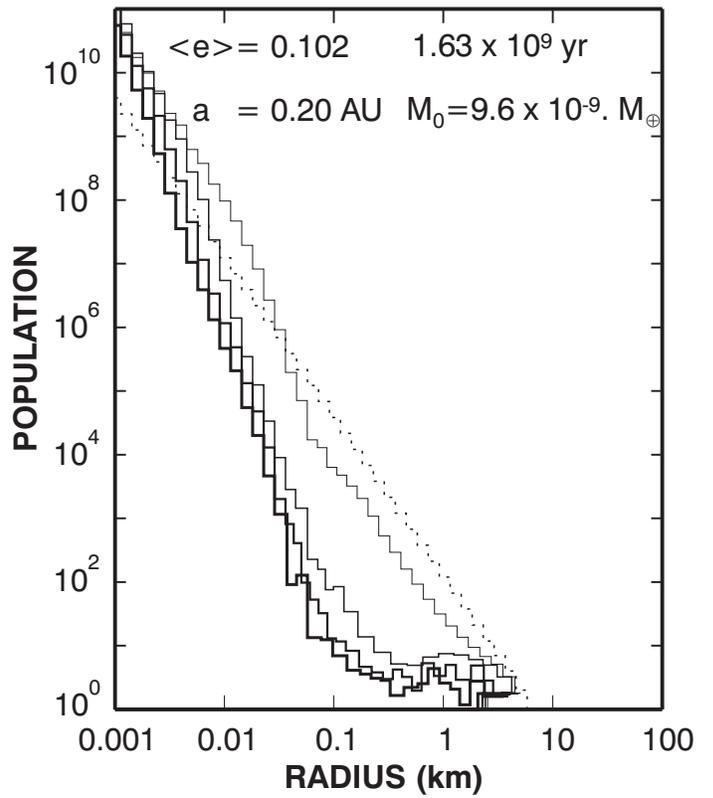
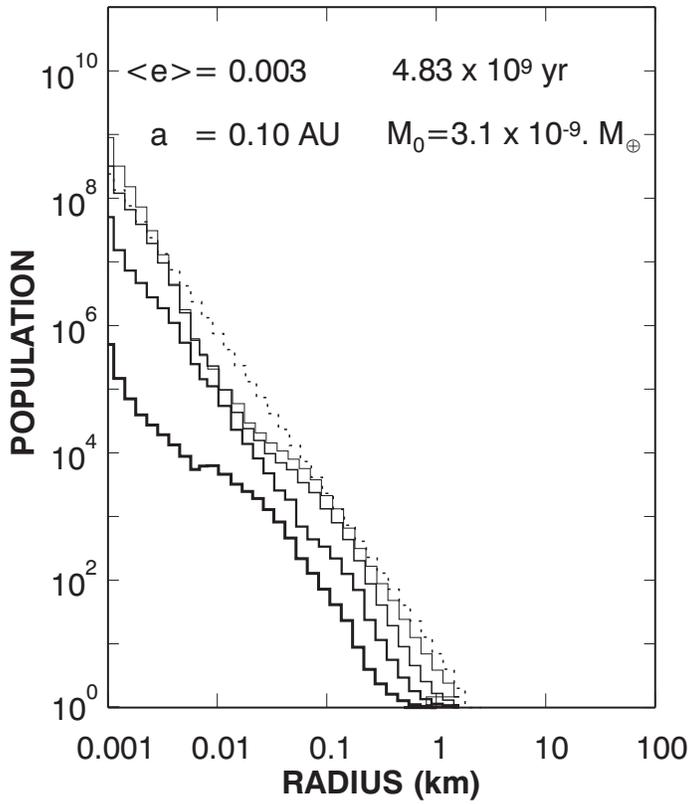
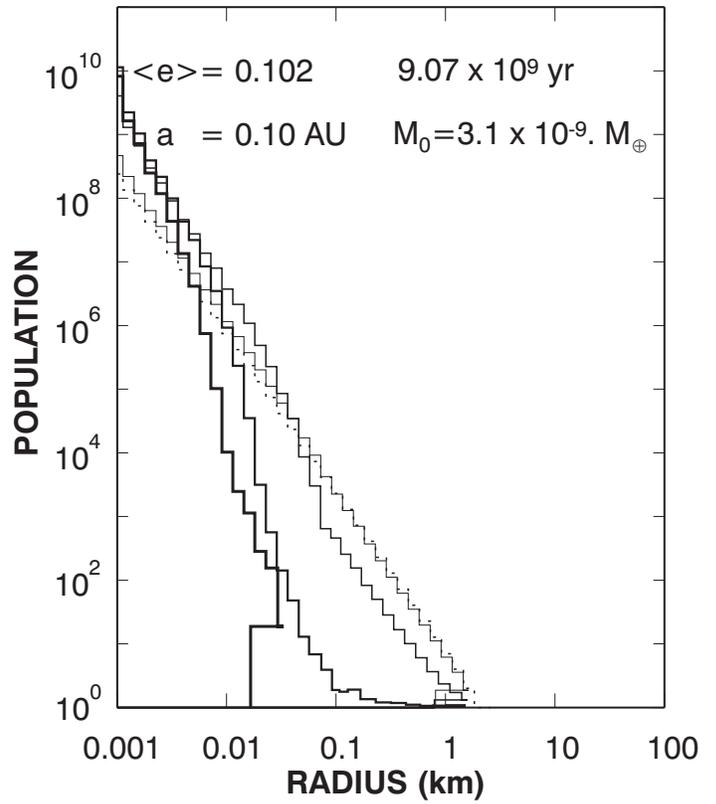
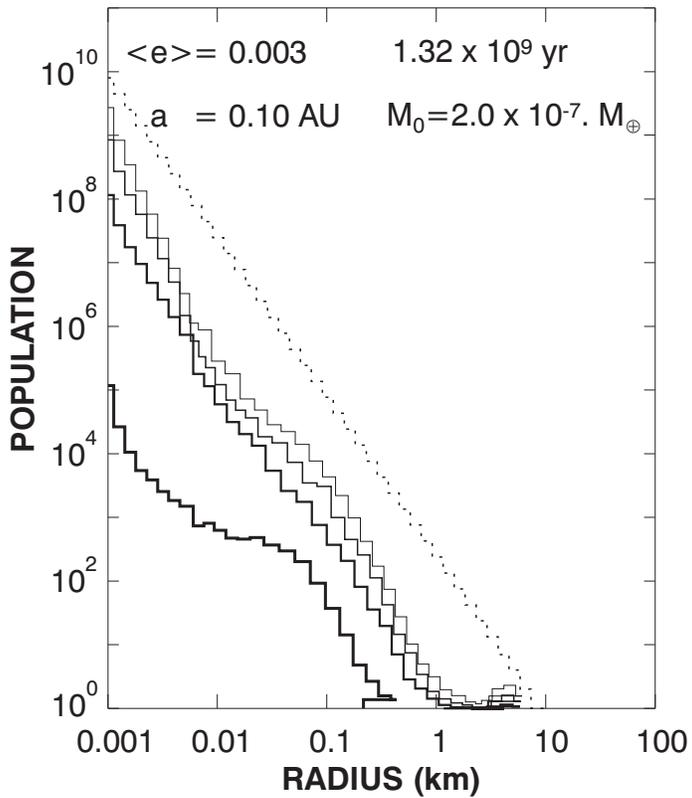
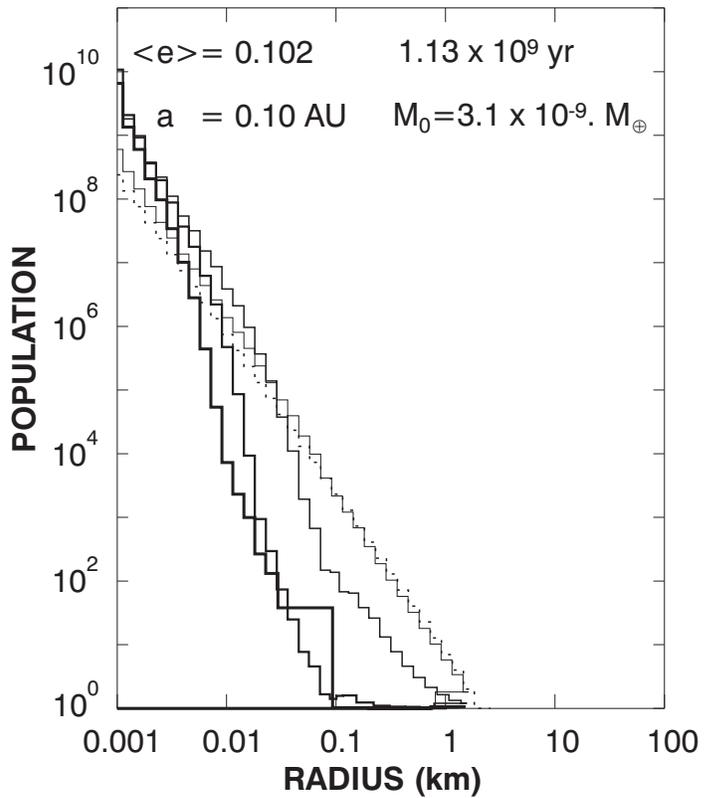
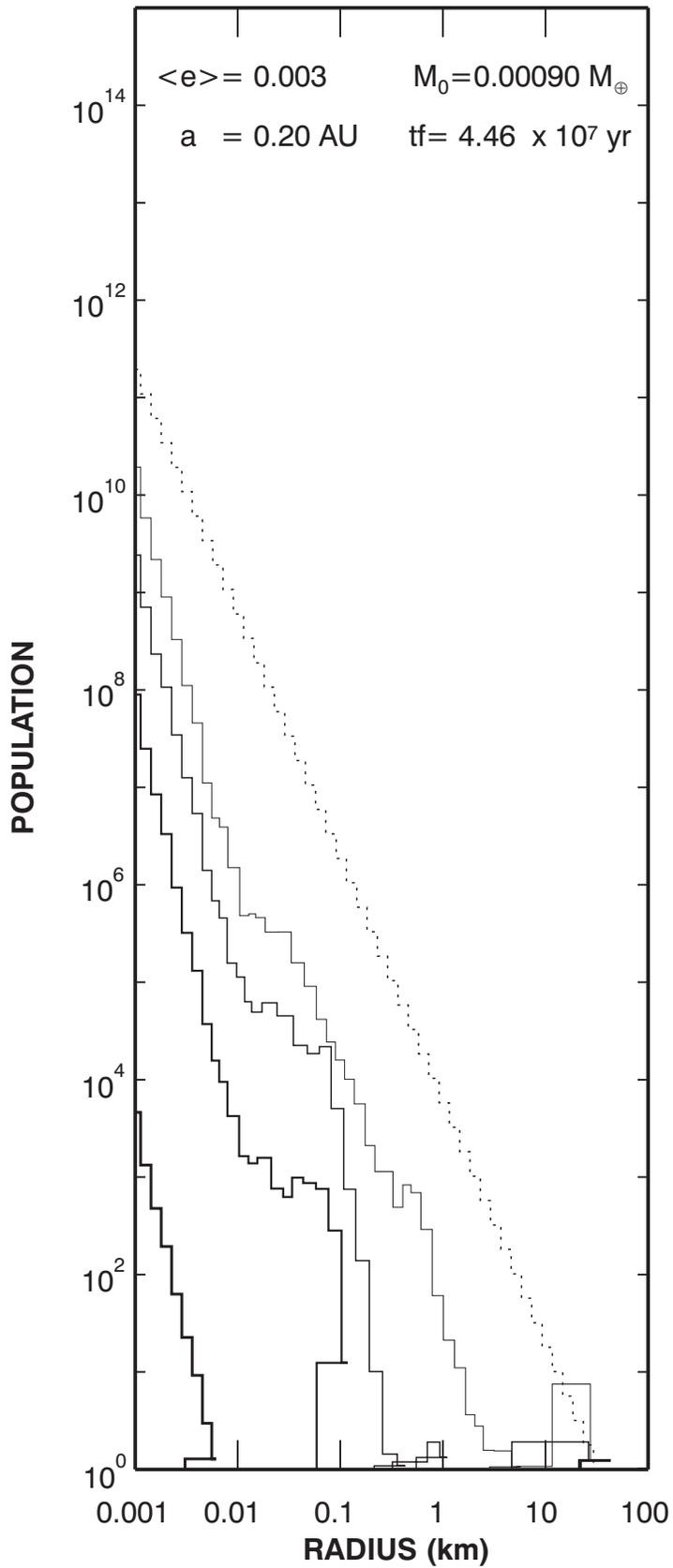
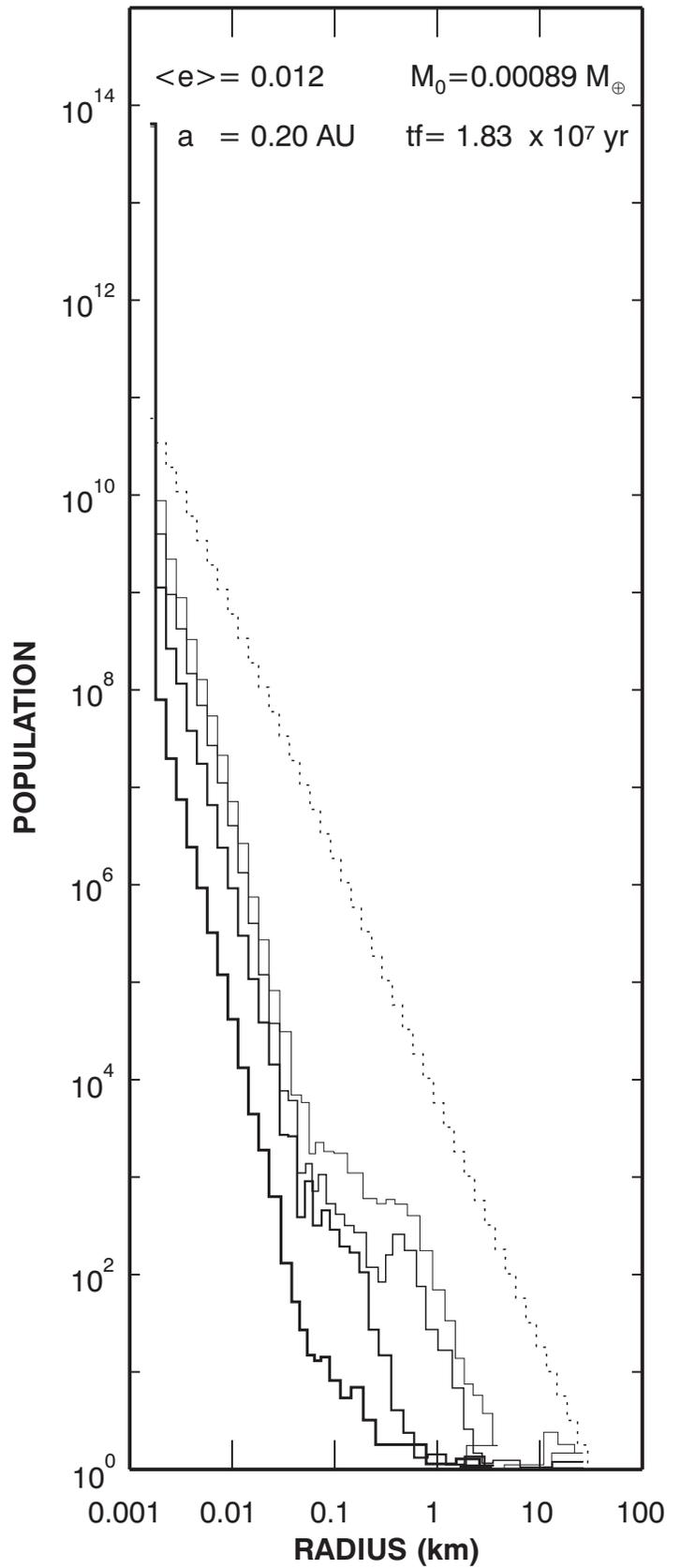
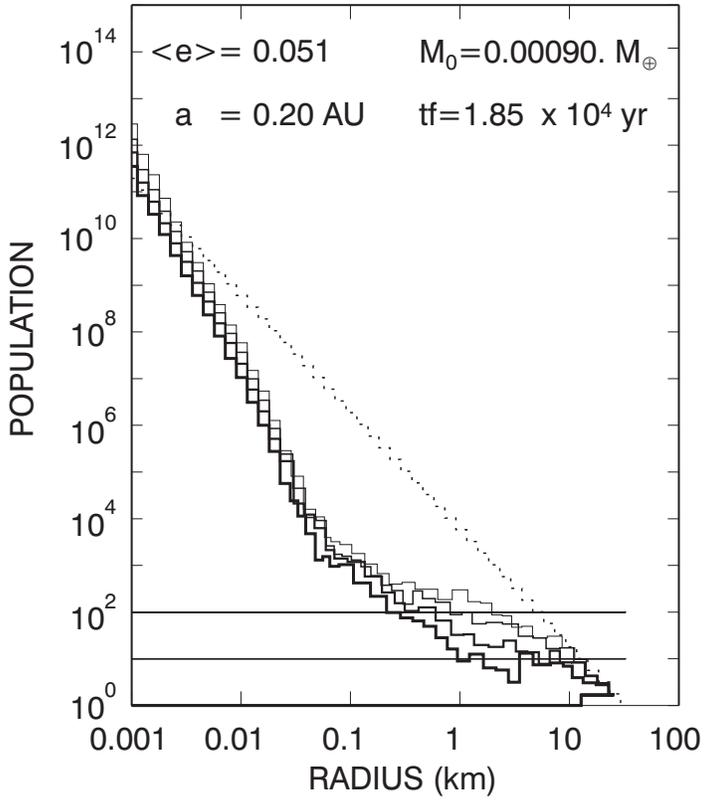
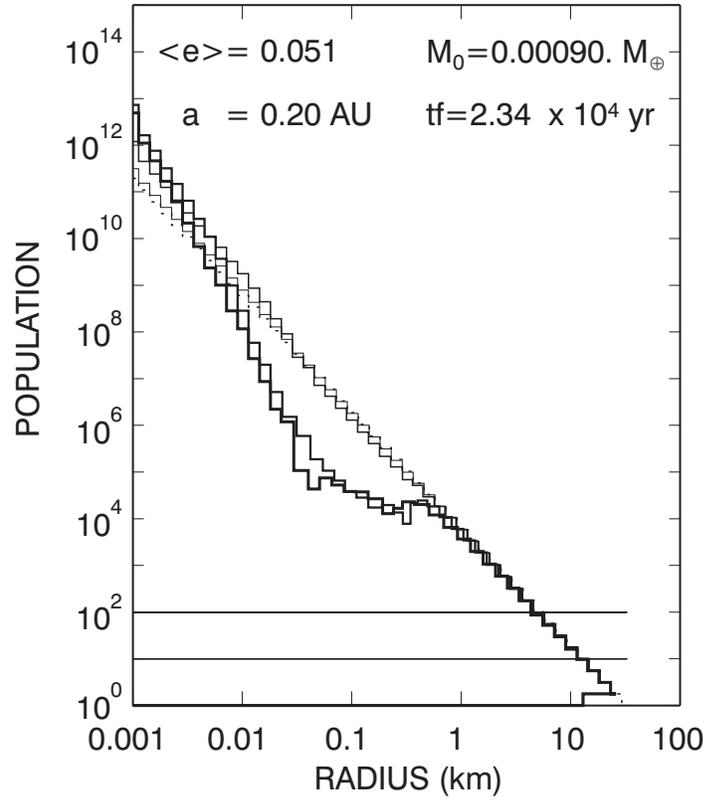
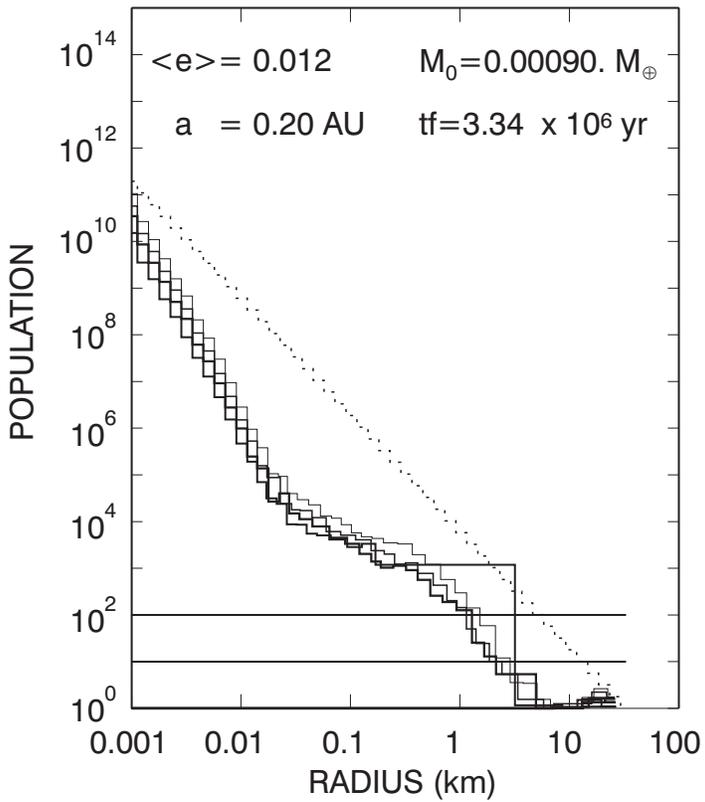
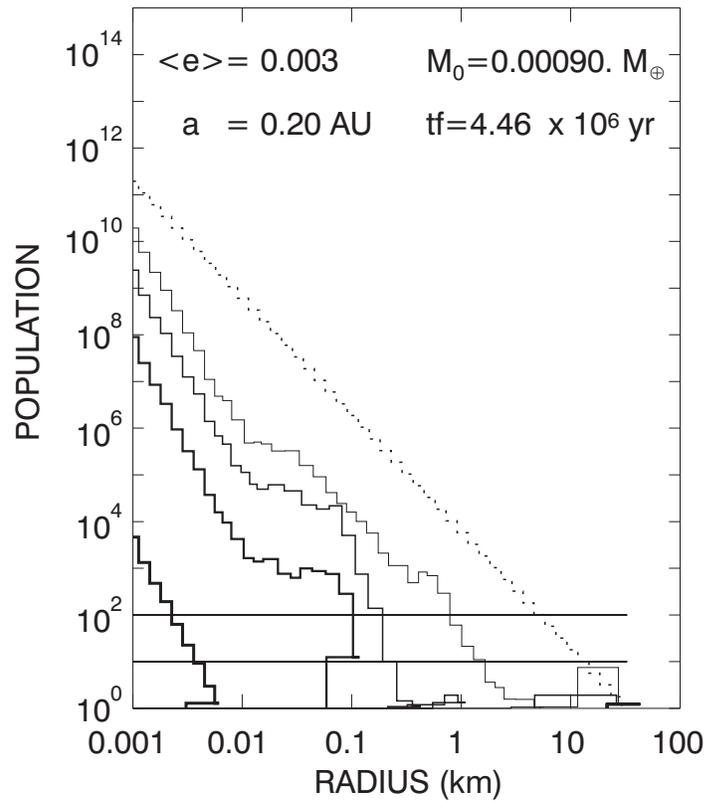
\bye